\documentclass[11pt]{article}
\usepackage{graphicx}
\usepackage{amsmath}

\begin{document}

\begin{titlepage}

\title{\bf Can the quintessence be a complex scalar field?\\
\vspace{0.2cm}}

\author{Je-An \ Gu\thanks{%
E-mail address: jagu@phys.ntu.edu.tw} \ \ \ and \ \ W-Y. P. Hwang\thanks{%
E-mail address: wyhwang@phys.ntu.edu.tw} \\
{\small Department of Physics, National Taiwan
University, Taipei 106, Taiwan, R.O.C.}
\medskip
}

\date{\small \today}

\maketitle

\begin{abstract}
In light of the recent observations of type Ia supernovae
suggesting an accelerating expansion of the Universe, we wish in
this paper to point out the possibility of using a complex scalar
field as the quintessence to account for the acceleration. In
particular, we extend the idea of Huterer and Turner in deriving
the reconstruction equations for the complex quintessence, showing
the feasibility of making use of a complex scalar field (instead
of a real scalar field) while maintaining the uniqueness feature
of the reconstruction for two possible situations respectively. We
discuss very briefly how future observations may help to
distinguish the different quintessence scenarios, including the
scenario with a positive cosmological constant.
\end{abstract}

\vspace{8.5cm}

\begin{flushleft}
\footnotesize \emph{PACS:} 98.80.Es, 98.80.Cq \\
\end{flushleft}

\end{titlepage}

\section{Introduction}
The Supernova Cosmology Project\footnote{http://snap.lbl.gov} and
the High-Z Supernova
Search\footnote{http://cfa-www.harvard.edu/cfa/oir/Research/supernova/HighZ.html}
reported on their observations of type Ia supernovae (SNe Ia),
suggesting that the expansion of the Universe is still
accelerating \cite{Perlmutter:1999np,Riess:1998cb}. In addition,
recent measurements of the power spectrum of the cosmic microwave
background (CMB) from BOOMERANG \cite{deBernardis:2000gy} and
MAXIMA-1 \cite{Hanany:2000} detected a sharp peak around $l \simeq
200$, indicating that the Universe is flat. Combining these two
classes of observations, we may conclude that the Universe has the
critical density (to make it flat) and that it consists of $1/3$
of ordinary matter and $2/3$ of dark energy with negative pressure
(such that $p_{total} < -\rho_{total} /3$ at present time)
\cite{deBernardis:2000cm,Balbi:2000tg}. At the moment, the most
often considered candidates for the dark energy include (1) the
existence of a positive cosmological constant \cite{Lambda models}
and (2) the presence of a slowly-evolving real scalar field called
"quintessence" \cite{Caldwell:1998ii}. In the case of the
quintessence, Zlatev, Wang and Steinhardt \cite{Tracker Q}
considered a real scalar field as the ``tracker field'' which,
regardless of a wide range of possible initial conditions, will
join a path more or less common to the evolving radiation, the
dominant energy density in the early universe, before the
matter-dominated era. In addition, Huterer and Turner
\cite{Huterer:1999ha} considered basically the inverse problem of
reconstructing the quintessence potential from the SNe Ia
observational data (see also Refs.\ \cite{Starobinsky:1998fr} and
\cite{Chiba:2000im}).

In this paper, we wish to point out the possibility of using a
complex scalar field as the quintessence to account for the
accelerating expansion of the Universe. In particular, we extend
the idea of Huterer and Turner in deriving the reconstruction
equations for the complex quintessence, showing the feasibility of
making use of a complex scalar field (instead of a real scalar
field) while maintaining the uniqueness of the solution of the
inverse problem. We also discuss briefly how future observations
may help to distinguish the different quintessence scenarios
(including the scenario with a positive cosmological constant).

\section{The Basics}
We consider a spatially flat Universe which is dominated by the
non-relativistic matter and a spatially homogeneous complex scalar
field $\Phi $ and which is described by the flat Robertson-Walker
metric:
\begin{equation}
ds^{2} = dt^{2} - a^{2}(t) \left( dr^{2} + r^{2}d\varphi _1^2 +
         r^{2}\sin ^2 \varphi _1 d\varphi _2^2 \right).
\label{metric tensor}
\end{equation}
The action for the Universe is
\begin{equation}
S = \int d^{4}x\sqrt{g} \left( -\frac{1}{16\pi G}\mathcal{R} -
\rho _{M} + \mathcal{L}_{\Phi} \right), %
\label{action}
\end{equation}
where $g$ is the absolute value of the determinant of the metric
tensor $g_{\mu \nu }$, $G$ is the Newton's constant, $\mathcal{R}$
is the Ricci scalar, $\rho _{M}$ is the matter density, and
$\mathcal{L}_{\Phi }$ is the Lagrangian density for the complex
scalar field $\Phi $:
\begin{equation}
\mathcal{L}_{\Phi } = \frac{1}{2}g^{\mu \nu}\left( \partial
        _{\mu}\Phi^{*} \right) \left( \partial _{\nu}\Phi \right)
        - V(\left| \Phi \right|), \quad \mu ,\nu = 0,1,2,3 .
\label{Lagrangian of Phi}
\end{equation}
In Eq.\ (\ref{Lagrangian of Phi}), we have assumed that the
potential $V$ depends only on the absolute value (or the
amplitude) of the complex scalar field: $\left| \Phi \right| $.

Instead of $\Phi $ and $\Phi ^{*}$, we would like to use the
alternative field variables: the amplitude $\phi (x)$ and the
phase $\theta (x)$ (of the complex scalar field $\Phi (x)$), which
are defined by
\begin{equation}
\Phi (x) = \phi (x) e^{i \theta (x)}. %
\label{Phi to phi&theta}
\end{equation}
(More precisely, $\Phi (t) = \phi (t) e^{i \theta (t)}$.) The
usage of the field variables---$\phi (x)$ and $\theta (x)$---will
benefit the derivation of the reconstruction equations which
relate the quintessence potential $V(\phi )$ to SNe Ia data. By
using Eq.\ (\ref{Phi to phi&theta}), the Lagrangian density for
$\Phi $ (Eq.\ (\ref{Lagrangian of Phi})) becomes
\begin{equation}
\mathcal{L}_{\Phi} = \frac{1}{2}g^{\mu \nu}\left( \partial
     _{\mu}\phi \right) \left( \partial _{\nu}\phi \right) +
     \frac{1}{2}\phi ^{2} g^{\mu \nu}\left( \partial
     _{\mu}\theta \right) \left( \partial _{\nu}\theta \right) -
     V(\phi ).
\end{equation}
The variation of the action (Eq.\ (\ref{action})) with the above
Lagrangian density yields the Einstein equations and the field
equations of the complex scalar field. By using the metric tensor
in Eq.\ (\ref{metric tensor}), these equations can be rearranged
and become
\begin{equation}
 H^2 \equiv \left( \frac{\dot{a}}{a} \right) ^{2} = %
   \frac{8\pi G}{3}\rho = \frac{8\pi G}{3} \left[ \rho _{M}
   + \left( \frac{1}{2} \dot{\phi}^{2}
   + \frac{1}{2} \phi ^{2} \dot{\theta}^{2} \right)
   + V(\phi ) \right] %
\label{Friedmann eqn}
\end{equation}
\begin{equation}
\left( \frac{\ddot{a}}{a} \right) = -\frac{4\pi G}{3} \left(
   \rho + 3p \right) = -\frac{8\pi G}{3} \left[ \frac{1}{2}\rho
   _{M} + \left( \dot{\phi}^{2} + \phi ^{2}\dot{\theta}^{2} \right)
   - V(\phi ) \right] %
\label{acceleration eqn}
\end{equation}
\begin{eqnarray}
\ddot{\phi} + 3H\dot{\phi}- \dot{\theta}^{2}\phi + V'(\phi )
&=& 0      \label{eqn of phi} \\
\ddot{\theta} + \left( 2 \frac{\dot{\phi}}{\phi} + 3H \right)
\dot{\theta} &=& 0 , \label{eqn of theta}
\end{eqnarray}
where $H$ is the Hubble parameter, dot and prime denote
derivatives with respect to $t$ and $\phi $ respectively, $\rho $
is the energy density, and $p$ is the pressure. We note that the
non-relativistic matter contributes the energy density $\rho _{M}$
and pressure $p_{M}=0$, while the evolving complex scalar field
contributes the energy density $\rho _{\Phi}$ and pressure
$p_{\Phi}$ as follows:
\begin{eqnarray}
\rho _{\Phi} &=& \frac{1}{2} \left( \dot{\phi}^{2} + \phi ^{2}
                 \dot{\theta}^{2} \right) + V(\phi ) \\
p_{\Phi}     &=& \frac{1}{2} \left( \dot{\phi}^{2} + \phi ^{2}
                 \dot{\theta}^{2} \right) - V(\phi ) .
\end{eqnarray}
Eqs.\ (\ref{Friedmann eqn})--(\ref{eqn of theta}) are the
fundamental equations which govern the evolution of the Universe.

We first note that Eq.\ (\ref{eqn of theta}) can be solved and the
solution for the ``angular velocity'' $\dot{\theta}$ is given by
\begin{equation}
\dot{\theta} = \frac{\omega}{a^{3}\phi ^{2}}, %
\label{soln of dot-theta}
\end{equation}
where $\omega $ is an integration constant determined by the
initial condition of $\dot{\theta}$ (or the value of
$\dot{\theta}$ at some specific time). By using Eq.\ (\ref{soln of
dot-theta}), the fundamental equations become
\begin{equation}
H^{2} \equiv \left( \frac{\dot{a}}{a} \right) ^{2} = %
   \frac{8\pi G}{3}\rho = \frac{8\pi G}{3} \left[ \rho _{M}
   + \frac{1}{2} \dot{\phi}^{2}
   + \frac{1}{2}\frac{\omega ^{2}}{a^{6}}\frac{1}{\phi ^{2}}
   + V(\phi ) \right] %
\label{Friedmann eqn 2}
\end{equation}
\begin{equation}
\left( \frac{\ddot{a}}{a} \right) = -\frac{4\pi G}{3} \left(
   \rho + 3p \right) = -\frac{8\pi G}{3} \left[ \frac{1}{2}\rho
   _{M} + \dot{\phi}^{2} + \frac{\omega ^{2}}{a^{6}}\frac{1}{\phi ^{2}}
   - V(\phi ) \right]  %
\label{acceleration eqn 2}
\end{equation}
\begin{equation}
\ddot{\phi} + 3H\dot{\phi} - \frac{\omega^2}{a^{6}}\frac{1}{\phi
^{3}} + V'(\phi) = 0.  %
\label{eqn of phi 2}
\end{equation}
Eq.\ (\ref{eqn of phi 2}) can be rearranged and become
\begin{equation}
\ddot{\phi} + 3H\dot{\phi} + \frac{d}{d\phi} \left[
\frac{1}{2}\frac{\omega ^{2}}{a^{6}}\frac{1}{\phi ^{2}} + V(\phi)
\right] = 0 .
\end{equation}
The term $\omega ^2 / (2a^6 \phi ^2)$ in the bracket, coming from
the ``angular motion'' of the complex scalar field $\Phi $, can be
treated as an effective potential (to be called ``centrifugal
potential''). It produces a ``centrifugal force'' and tends to
drive $\phi $ away from zero if $\omega \neq 0$ (i.e.\ the
``angular velocity'' $\dot{\theta}$ is nonzero).

We note that, because the gravitational influence of the dark
energy with negative pressure tends to prevent the ordinary matter
from forming structures, the contributions from quintessence to
the energy density and pressure should have been insignificant
just a short time ago. Furthermore, for preserving the concordance
between theories (Big Bang Cosmology + Inflation + Cold Dark
Matter) and observations (for instance, measurements of CMB and
light element abundances), the contributions from quintessence
should also be negligible at the epochs of `recombination' and
`primordial nucleosynthesis'. This constraint on quintessence will
help to disentangle two kinds of kinetic energy provided by the
complex quintessence, one from evolving $\phi$ and the other from
the ``angular motion'' of the complex scalar field $\Phi $, as
follows.

In Eqs.\ (\ref{Friedmann eqn 2}) and (\ref{acceleration eqn 2}),
the contributions from the ``angular motion'' of $\Phi $ to the
energy density $\rho $ and pressure $p$ both are proportional to
$a^{-6}\phi ^{-2}$. The factor $a^{-6}$ may make these
contributions decrease very fast, even faster than the matter
density $\rho _{M}$ (which is proportional to $a^{-3}$), provided
that $\phi $ does not decrease as fast as $a^{-3/2}$ (i.e.\ along
with the expansion of the Universe). In this situation, the
angular-motion contributions are negligible at the present epoch,
since they should have been insignificant just a short time ago.
On the other hand, under the situation that $\phi $ decreases no
slower than $a^{-3/2}$, the angular-motion contributions to the
energy density and pressure are no longer necessary to be
negligible, while the contributions from evolving $\phi $
(proportional to $\dot{\phi}^{2}$) in Eqs.\ (\ref{Friedmann eqn
2}) and (\ref{acceleration eqn 2}) would decrease no slower than
$a^{-3}H^{2}$. In this situation, the evolving-$\phi $
contributions can be neglected at the present epoch, since they
should also have been insignificant a short time ago.

To sum up, we have two possible situations, depending on how fast
$a^{-6}\phi^{-2}$ falls off, compared with $\rho_M$ (i.e.\
$a^{-3}$), as the Universe expands. One situation is that the
contributions from the ``angular motion'' of the complex scalar
field $\Phi $ to the energy density and pressure are negligible.
The other is that the contributions from evolving $\phi $ are
negligible. Indeed the disentanglement of these two types of
kinetic energy in these two situations plays a crucial role in
making the reconstruction of the quintessence potential feasible,
as will be discussed in the next section. In addition, we note
that these two possible situations are corresponding to two kinds
of quintessence. The one with negligible angular-motion
contributions behaves like the real quintessence. The other with
negligible evolving-$\phi$ contributions may be distinct from the
real quintessence if the angular-motion contributions are
significant. When both the contributions from evolving $\phi $ and
the ``angular motion'' of the complex scalar field $\Phi $ are
negligible such that the quintessence potential dictates, it is
equivalent to the case of having a positive cosmological constant.

\section{Reconstruction Equations}
The quantities introduced in Sec.\ 2
\begin{equation}
\begin{array}{ccl}
t    &:&  \mbox{Robertson-Walker time coordinate} \\
a(t) &:&  \mbox{scale factor}  \\
H(t) &:&  \mbox{Hubble parameter}  \\
\rho _{M}(t) &:&  \mbox{matter energy density}
\end{array}
\label{non-observables}
\end{equation}
are neither observable quantities themselves in experiments, nor
directly related to observations. In order to obtain the
reconstruction equations for the quintessence potential $V(\phi
)$, we need to use the observationally relevant quantities
\begin{equation}
\begin{array}{ccl}
z      &:&  \mbox{redshift} \\
r(z)   &:&  \mbox{Robertson-Walker coordinate distance} \\
       & &  \mbox{to an object at redshift z} \\
H_{0}  &:&  \mbox{Hubble constant} \\
\Omega _{M} &:&  \mbox{matter energy density fraction}
\end{array}
\label{observables}
\end{equation}
to replace them. The quantities in (\ref{non-observables}) and
(\ref{observables}) are related by
\begin{eqnarray}
1+z  &=& \frac{1}{a}  \label{redshift} \\
r(z) &=& -\int _{t_0}^{t(z)}\frac{dt'}{a(t')}=\int _0^z
         \frac{dz'}{H(z')}  \label{r of z} \\
H(z) &=& \frac{\dot{a}}{a} = \frac{1}{(dr/dz)}  \label{Hubble parameter} \\
\rho _{M} &=& \Omega _{M} \rho _{c} = \frac{3\Omega
              _{M}H_{0}^{2}}{8\pi G}(1+z)^{3} ,  \label{matter density}
\end{eqnarray}
where $\rho _{c}$ is the critical density, and we have set the
present scale factor $a_{0}$ to be one. In addition, for the
quantity $\ddot{a}/a$ in Eq.\ (\ref{acceleration eqn 2}), which is
related to the acceleration of the expansion, we have
\begin{equation}
\frac{\ddot{a}}{a} =  \frac{1}{(dr/dz)^{2}} +
                      (1+z)\frac{d^{2}r/dz^{2}}{(dr/dz)^{3}}.
\label{acceleration vs r&z}
\end{equation}

Using Eqs.\ (\ref{redshift})--(\ref{acceleration vs r&z}),
we can obtain the reconstruction equations from the fundamental
equations (\ref{Friedmann eqn 2}) and (\ref{acceleration eqn 2}).
These reconstruction equations relate $V(\phi )$, $\dot{\phi }$,
and $\dot{\theta }$ to the observationally relevant quantities
$z$, $r(z)$, $H_0$, and $\Omega _M$ as follows:
\begin{equation}
V \left[ \phi (z) \right] = \frac{1}{8\pi G} \left[
\frac{3}{(dr/dz)^2}+(1+z)\frac{d^2 r/dz^2}{(dr/dz)^3} \right] -
\frac{3\Omega _M H_0^2}{16\pi G}(1+z)^3 ,  %
\label{reconstruct V}
\end{equation}
\begin{equation}
\left( \frac{d\phi}{dz} \right) ^2 + \frac{\omega ^2}{\phi
^2}(1+z)^4 \left( \frac{dr}{dz} \right) ^2 =
\frac{(dr/dz)^2}{(1+z)^2} \left[ -\frac{1}{4\pi G} \frac{(1+z)(d^2
r/dz^2)}{(dr/dz)^3} - \frac{3\Omega _M H_0^2}{8\pi G}(1+z)^3
\right] ,  %
\label{reconstruct phi}
\end{equation}
where the term $(\omega ^2 / \phi ^2) (1+z)^4 (dr/dz)^2$ in Eq.\
(\ref{reconstruct phi}) is the contribution from the ``angular
motion'' of the complex scalar field $\Phi $.

Through the reconstruction equations, given the data $r(z)$ from
SNe Ia experiments and the values of the parameters $\Omega _M$
and $H_0$ obtained from other experiments, it seems that we can
reconstruct the quintessence potential $V(\phi )$ after inputting
the values of $\omega $ and $\phi _0$, which correspond to some
specific initial conditions. But, unlike the case of a real scalar
field discussed in \cite{Huterer:1999ha}, the values of $\omega $
and $\phi _0$ are not the parameters we can input arbitrarily.
They will determine the proportion of the contributions from the
``angular motion'' of the complex scalar field $\Phi $ to the
energy density and pressure. As discussed in Sec.\ 2, we have two
possible situations, the one in which the contributions from the
``angular motion'' of the complex scalar field $\Phi $ are
negligible and the other one in which the contributions from
evolving $\phi $ are negligible. In the following, however, we
will show that the reconstruction of $V(\phi )$ is still possible
for both situations.

For the situation in which the angular-motion contributions are
negligible, i.e.\ $\phi$ does not decrease as fast as $a^{-3/2}$,
the reconstruction equations become the same as those in
\cite{Huterer:1999ha} as follows:
\begin{eqnarray}
V \left[ \phi (z) \right] &=& \frac{1}{8\pi G} \left[
       \frac{3}{(dr/dz)^2}+(1+z)\frac{d^2 r/dz^2}{(dr/dz)^3}
       \right] - \frac{3\Omega _M H_0^2}{16\pi G}(1+z)^3
       \label{reconstruct V for case 1} \\
\left( \frac{d\phi}{dz} \right) ^2 &=&
       \frac{(dr/dz)^2}{(1+z)^2} \left[ -\frac{1}{4\pi G}
       \frac{(1+z)(d^2 r/dz^2)}{(dr/dz)^3} -
       \frac{3\Omega _M H_0^2}{8\pi G}(1+z)^3 \right] .
       \label{reconstruct phi for case 1}
\end{eqnarray}
We note that in this case only the shape of $V(\phi )$ and the
corresponding region of $V(\phi )$ (which corresponds to the
observational region of redshift $z$) can influence the evolution
of the Universe, while the value of $\phi _0$ has no influence.
Thus, in this case, the initial value $\phi _0$ can be put in by
hand, and $\omega $ is no longer the parameter we need to input.
With the initial value $\phi _0$, the parameters: $\Omega _M$ and
$H_0$, and the data $r(z)$, we can obtain the information on
$V(z)$ and $\phi (z)$ through the reconstruction equations
(\ref{reconstruct V for case 1}) and (\ref{reconstruct phi for
case 1}), respectively, and then reconstruct the quintessence
potential $V(\phi )$ for some specific region of $V(\phi )$
corresponding to the observational region of redshift $z$. We note
that the reconstruction of $V(\phi )$ in this case is in the same
way as the case of a real scalar field discussed in
\cite{Huterer:1999ha}.

On the other hand, for the situation in which the evolving-$\phi $
contributions are negligible, i.e.\ $\phi$ decreases no slower
than $a^{-3/2}$, the reconstruction equations become
\begin{eqnarray}
V \left[ \phi (z) \right] &=& \frac{1}{8\pi G} \left[
       \frac{3}{(dr/dz)^2}+(1+z)\frac{d^2 r/dz^2}{(dr/dz)^3}
       \right] - \frac{3\Omega _M H_0^2}{16\pi G}(1+z)^3
       \label{reconstruct V for case 2} \\
\left( \frac{\omega }{\phi } \right) ^2  &=& \frac{1}{(1+z)^6}
\left[ -\frac{1}{4\pi G}
       \frac{(1+z)(d^2 r/dz^2)}{(dr/dz)^3} -
       \frac{3\Omega _M H_0^2}{8\pi G}(1+z)^3 \right] .
       \label{reconstruct phi for case 2}
\end{eqnarray}
With the initial value $\phi _0$, the parameters: $\Omega _M$ and
$H_0$, and the data $r(z)$, Eq.\ (\ref{reconstruct phi for case
2}) may be used to determine the value of $\omega $ and yield the
information on $\phi (z)$, and Eq.\ (\ref{reconstruct V for case
2}) gives the information on $V(z)$ in the observational region of
redshift $z$. Then the reconstruction of the quintessence
potential $V(\phi )$ can be achieved for some region of $V(\phi )$
corresponding to the observational region of redshift $z$. Unlike
the previous case, the reconstruction of $V(\phi )$ in this case
is different from the case of a real scalar field discussed in
\cite{Huterer:1999ha}, because the possible existence of a
significant angular-motion contribution here is the very feature
that the real quintessence does not possess.

As shown in the above, two viable quintessence potentials may be
reconstructed, with the use of SNe Ia data, for two possible
situations respectively. Nevertheless, the self-consistency of the
reconstruction through Eqs.\ (\ref{reconstruct V for case 2}) and
(\ref{reconstruct phi for case 2}), for the situation in which the
evolving-$\phi$ contributions are negligible, should be checked,
as follows. From Eq.\ (\ref{reconstruct phi for case 2}), we can
obtain $\phi (z)$, and also $\phi (a)$. We then check whether
$\phi (a)$ fits the requirement that the evolving-$\phi$
contributions (proportional to $\dot{\phi}^2$) are negligible at
the present epoch. If $\phi (a)$ does not pass this consistency
check, the complex quintessence model with significant
angular-motion contributions is ruled out, and we can only use
Eqs.\ (\ref{reconstruct V for case 1}) and (\ref{reconstruct phi
for case 1}) to reconstruct the quintessence potential instead.
Likewise, for the situation in which the angular-motion
contributions are negligible, we can check whether $\phi (a)$,
obtained from Eq.\ (\ref{reconstruct phi for case 1}), fits the
`sufficient condition' of this situation: $\phi$ does not decrease
as fast as $a^{-3/2}$. If not, the reconstruction of the
quintessence potential through Eqs.\ (\ref{reconstruct V for case
1}) and (\ref{reconstruct phi for case 1}) will be accompanied by
a delicate choice of a small enough ``angular-velocity'' parameter
$\omega$, therefore possessing a naturalness problem.


\section{Discussion and Summary}
In this work, we have investigated the scenario of using a complex
scalar field as the quintessence for accelerating the expansion of
the Universe. In the present scenario, there are two kinds of
``kinetic-energy'' type contributions to the energy density and
pressure, one coming from the evolving amplitude $\phi $ and the
other from the ``angular motion'' of the complex scalar field
$\Phi $. In many cases, the contribution from the ``angular
motion'' may decrease very fast along with the expansion of the
Universe, and is negligible in the process of reconstructing the
quintessence potential $V(\phi )$ (which is responsible for the
possible accelerating expansion of the Universe). Nevertheless,
there is also a situation in which the contribution from the
evolving amplitude $\phi $ is negligible while the part from the
``angular motion'' need to be treated with care.

Making use of the reconstruction equations (\ref{reconstruct V})
and (\ref{reconstruct phi}) (as derived from the fundamental
equations
(\ref{Friedmann eqn})--
(\ref{eqn of theta})), we may reconstruct the quintessence
potential $V(\phi )$ from the observational data $r(z)$ (the
coordinate distance as a function of the redshift $z$, as may be
deduced from SNe Ia experiments), respectively for the two
situations mentioned above. Accordingly, the complex scalar fields
may be used as the candidate for the quintessence and our analysis
indicates that, depending on how fast $a^{-6}\phi^{-2}$ falls off
as the Universe expands, the quintessence potential may be
reconstructed in a fairly unique manner.

It is useful to note that the observation data on $r(z)$ may be
converted uniquely into the information on the effective equation
of state of the dark energy: $w(z) \equiv p_X(z) / \rho_X(z)$
(where `X' denotes the dark energy)
\cite{Chiba:2000im,Nakamura:1999mt}. Such information may in turn
be used to distinguish different quintessence scenarios: The
scenario with a positive cosmological constant corresponds to
$w=-1$ so that a significant variation of $w$, especially
differing from the value of $-1$, would help to rule out such
scenario. The distinction between the real and complex scalar
field scenarios is obviously more subtle: When the ``angular
motion'' part is negligible, the complex quintessence behaves like
the real one. However, the situation when the ``angular motion''
part is more important by comparison needs to be further studied
since it poses new possibilities for the Universe.

In any event, the complex scalar fields as the quintessence should
be seriously considered since such fields, unlike the real scalar
field, have been invoked in many different sectors of elementary
particle physics, such as the possibility for gauging (i.e.,
interacting with the various gauge fields), responsible for mass
generations (Higgs mechanisms), as well as arising from
condensates into Goldstone ``pions'' (from techni-color
quark-antiquark pairs). In our opinion, the rich physics
associated with the complex fields should be taken seriously in
constructing a workable model, or theory, for the early universe.

\section*{Acknowledgements}
This work is supported in part by the National Science Council,
Taiwan, R.O.C. (NSC 89-2112-M-002-062) and by the CosPA project of
the Ministry of Education (MOE 89-N-FA01-1-4-3).

\end{document}